\documentclass[useAMS,usenatbib]{mnras}
\usepackage{epsf}
\usepackage{amssymb}
\usepackage{epsfig}
\usepackage[usenames]{color}
\usepackage{float}
\usepackage{soul}
\usepackage{xcolor}
\usepackage{xspace}

\definecolor{orange}{RGB}{255,127,0}
\definecolor{dgreen}{RGB}{1,70,28}

\newcommand*\src{4U\,1543$-$624~}
\newcommand*\srcs{Swift\,J1756.9$-$2508~}

\newcommand{\nicer}{\textit{NICER}\xspace}

\newcommand{\xmm}{{\it XMM-Newton}\xspace}
\newcommand{\nus}{{\it NuSTAR}\xspace}

\newcommand{\cxo}{\hbox{Chandra}\xspace}

\title[Iron K$\alpha$ line variability in UCXBs]{Disappearance of the Fe K$\alpha$ emission line in Ultra Compact X-ray Binaries {\src} and  \srcs}
\author[F.Koliopanos et al.]{Filippos Koliopanos$^{1,2}$\thanks{fkoliopanos@irap.omp.eu}, Georgios Vasilopoulos$^3$, Sebastien Guillot$^{1,2,4}$ and
\newauthor 
Natalie Webb$^{1,2}$ \\
$^{1}$Universit{\'e} de Toulouse; UPS-OMP; IRAP, Toulouse, France\\
$^{2}$CNRS, IRAP, 9 Av. colonel Roche, BP 44346, F-31028 Toulouse cedex 4, France\\
$^{3}$Department of Astronomy, Yale University, PO Box 208101, New Haven, CT 06520-8101, USA\\
$^{4}$CNES, IRAP, 9 Av. colonel Roche, BP 44346, F-31028 Toulouse cedex 4, France\\}

\begin{document}

\date{Accepted .... Received ...}

\pagerange{\pageref{firstpage}--\pageref{lastpage}} \pubyear{2020}

\maketitle

\label{firstpage}

\begin{abstract}
We investigate the long-term variability of the iron K${\alpha}$ line in the spectra of two Ultra Compact X-ray Sources (UCXBs) with C/O-rich donors. We revisit archival observations from five different X-ray telescopes, over a ${\sim}$twenty year period. Adopting physically motivated models for the spectral continuum, we probe the long-term evolution of the source emission in a self-consistent manner enabling physical interpretation of potential variability of the primary X-ray continuum emission and/or any emission lines from reflection off the accretion disk. We find that the spectral shape and flux of the source emission (for both objects) has remained almost constant throughout all the observations, displaying only minor variability in some spectral parameters and the source flux (largest variation is a ${\sim}$25\% drop in the flux of \srcs). We note a striking variability of the Fe K${\alpha}$ line which fluctuates from a notable equivalent width of ${\sim}$66-100\,eV in \src and ${\sim}$170\,eV in \srcs, to non-detections with upper limits of 2-8\,eV. We argue that the disappearance of the iron line is due to the screening of the Fe K${\alpha}$ line by the overabundant oxygen in the C/O-rich UCXBs. This effect is cancelled when oxygen becomes fully ionized in the inner disk region, resulting in the variability of the Fe K${\alpha}$ line in an otherwise unaltered spectral shape. This finding supports earlier predictions on the consequences of H-poor, C/O-rich accretion disk on reflection induced fluorescent lines in the spectra of UCXBs.
\end{abstract}


\section{Introduction}
Ultra compact X-ray binaries (UCXBs) are accreting binary systems, defined by their very short (less than 1\,hr) orbital periods. Their periods suggest such tight orbits that a main sequence star cannot fit  (e.g., \citealt{1984ApJ...283..232R}; \citealt*{1986ApJ...311..226N}). 
Evolutionary scenarios and observational findings indicate that UCXBs are composed of a Roche lobe filling white dwarf or helium star that is accreting  material on to a neutron star \citep[e.g.,][]{1993ARep...37..411T, 1995ApJS..100..233I, 1995xrbi.nasa..457V, 2003ApJ...598.1217D, 2005ApJ...624..934D}.
{ More specifically, the different formation paths of UCXBs -- which may or may not include a common envelope phase with their binary companion -- result in donor stars that can range from He stars or He-WDs to C/O or O/Ne/Mg-WDs~\citep[e.g.,][]{1986A&A...155...51S,2002ApJ...565.1107P,2002A&A...388..546Y, 2004ApJ...607L.119B}. For simplicity UCXBs are often divided into  two main categories: He-rich and He-poor, depending on the donor composition.} Since divergent UCXB formation paths lead to degenerate donors of similar mass, determining the chemical composition of the disk (and therefore the donor star) in these sources is a crucial step towards constraining their formation history. Furthermore, UCXBs offer a unique opportunity to study accretion of hydrogen-poor matter onto compact objects. 

The non-solar abundance of the accreting material in UCXBs can have a profound effect on the emission line spectrum of UCXBs in both the optical \citep[He, C, or O lines, e.g.,][]{2004MNRAS.348L...7N, 2006A&A...450..725W,2006MNRAS.370..255N} and the X-ray wavelengths (primarily in the form of C and O K${\alpha}$ lines, e.g.,~\citealt{2001ApJ...560L..59J,2005ApJ...627..926J,2010MNRAS.407L..11M}). Nevertheless, due to the very faint optical counterparts of UCXBs (V-band absolute magnitudes ${\gtrsim}5$) but also due to the increased interstellar absorption in the $<1$\,keV range of the X-ray spectrum the detection of C and O emission lines is often considerably difficult. In a relatively recent theoretical study, \cite{2013MNRAS.432.1264K} demonstrated that the iron $\rm K{\rm\alpha}$ line located at 6.4\,keV (and therefore not affected by interstellar absorption) can be used as an indirect method for determining the chemical composition of the accretion disk and donor star in UCXBs.

More specifically, for moderately luminous sources (${\rm L}\rm_X\lesssim$ a few $10^{37}\rm erg\,s^{-1}$)  a strong suppression of the Fe $\rm K{\rm\alpha}$ line was predicted in the case of a C/O or O/Ne/Mg WD donor, translating to a more than tenfold decrease of the equivalent width (EW) of the line. On the other hand, in the case of He-rich disks, the iron line remains unaffected with an EW in the same range as ``standard'' X-ray binaries (XRBs) with hydrogen rich donors. Observational analysis of five well-known UCXB sources corroborated these predictions \citep{2014MNRAS.442.2817K}. In addition to the above predictions, \cite{2013MNRAS.432.1264K} indicated that the screening effect is decidedly linked to the ionization state of oxygen and (to a second degree) carbon. The authors further demonstrated that -- given some specific assumptions -- the ionization state of C and O in the disk and subsequently the presence or absence of the prominent Fe $\rm K{\rm\alpha}$ line is luminosity dependent. In this paper we report on the behavior of two known UCXBs, which appear to exhibit the iron line variability, predicted by \cite{2013MNRAS.432.1264K}.

{\src} is a well known UCXB which most likely hosts a C/O or Ne-rich WD donor \citep[e.g.,][]{2001ApJ...560L..59J}. Discovered by the UHURU telescope \citep{1977ApJ...214..856J}, it was classified as an UCXB after \cite{2004ApJ...616L.139W} established a period of P${=}$18.2 minutes, based on optical light curves. \src is a persistent X-ray source with a stable, moderately bright emission since its discovery. It has been observed by most major X-ray observatories, including BEPPOSAX \citep{2003A&A...402.1021F}, ASCA and RXTE \citep{2003A&A...397..249S} as well as Chandra and {\xmm} \citep{2001ApJ...560L..59J,2003ApJ...599..498J,2011MNRAS.412L..11M,2014MNRAS.442.1157M} and more recently NICER (this paper and \citealt{2019ApJ...883...39L}). The  X-ray continuum of the source has been modeled using a variety of different models including thermal and non-thermal components leading to different interpretations for its physical origin. Furthermore, a pronounced emission line centered at ${\sim}0.7$\,keV  has been detected by both the {\xmm} RGS and the Chandra HETG spectrometers. The emission line has been attributed to reflection of hard X-rays from a C/O-rich disk \citep[most prominent being the O {\small VII} Ly${\alpha}$ line, e.g., ][]{2003ApJ...599..498J,2011MNRAS.412L..11M,2014MNRAS.442.1157M}. The presence of the iron K${\alpha}$ line -- usually detected in the spectra of X-ray binaries -- has been much more dubious. The line was reported in BEPPOSAX and RXTE but was not present in ASCA spectra \citep{2001ApJ...560L..59J,2003A&A...397..249S}. Its potential presence has been tentatively claimed in the {\xmm} spectra \citep{2011MNRAS.412L..11M,2014MNRAS.442.1157M} but not in Chandra \citep{2003ApJ...599..498J}. { More recently, \cite{2019ApJ...883...39L} presented a thorough analysis of X-ray and radio observations of \src during an enhanced accretion episode in 2017 \citep{2017ATel10690....1L,2017ATel10719....1M}. The iron emission line re-emerged during this phase, \cite{2019ApJ...883...39L} reported its presence and attributed it to X-ray reflection off the accretion disk surface.}

\srcs is an LMXB located in the direction of the Galactic bulge. It was discovered in 2007 during an X-ray outburst observed by the Swift-BAT. Follow up observations with RXTE identified the source as an UCXB ($P_{orb}{\sim}54.7$\,min) with a millisecond pulsar \citep{2007ApJ...668L.147K}. The transient source was again detected by the Swift-BAT and RXTE
PCA during a second outburst in 2009 and more recently in 2018, when it was observed by Swift/XRT, \xmm, {\nus} and NICER \citep{2018ATel11603....1K,2018ATel11502....1B}. The multiple observations spanning more than a decade allowed the accurate study of the orbital evolution of the source \citep{2010MNRAS.403.1426P,2018MNRAS.481.1658S,2018ApJ...864...14B}, which in turn indicated the strength of the NS magnetic field at ${\sim}2{\times}10^{8}$\,G \citep{2018MNRAS.481.1658S}. No type-I X-ray bursts were detected during any of the outbursts of \srcs.

In this paper we revisit observations of \src and \srcs at different epochs, focusing our analysis on the shape of the spectral continuum and the presence or absence of the iron emission line. More specifically we examine observations of \src between 1997 and 2001 as well as its unusually bright 2017 outburst \citep[\src is a persistent source, which entered a remarkably bright phase on Sept. 2017, ][]{2017ATel10719....1M}. For the transient source \srcs we study and compare its two outbursts of 2009 and 2018. We demonstrate that the X-ray spectrum of both sources has more or less the same shape through the years, which can be modeled using simple but physically motivated models. More importantly, we show that the flux of the Fe K${\alpha}$ emission detected on top of the continuum is variable by more than an order of magnitude. We discuss this behavior in the context of our 2013 prediction for Fe line variability in C/O-rich UCXBs and utilize our findings to further scrutinize and extend our initial hypothesis.

\section{Observations, data analysis and results}
\label{sec:data}

In this work we revisit archival 1997 RXTE, 2000 Chandra High Energy Transmission Grating (HETG), 2001 {\xmm} and 2017 NICER observations of {\src} and 2009 RXTE and 2017 {\it XMM-Newton/NuSTAR} observations of \srcs. Details of the observations are tabulated in Table~\ref{tab:obs2}. We note that in the case of the RXTE and NICER observations there are multiple pointings within a period of several days during the source outburst. We have reviewed all available observations and selected those presented Table~\ref{tab:obs2} based on their high number of registered counts. We have verified that within all data sets, the source was observed in the same spectral state. Analysis of the source spectra was carried out using the {\small{XSPEC}} X-ray spectral fitting package, Version 12.9.1 \citep{1996ASPC..101...17A}.

\subsection{Modeling the X-ray spectral continuum}
\label{sec:hard}
{ In this analysis we use a combination of multi-color disk black body and nominal black body spectral components to model the X-ray emission continuum of our two sources. Our choice of the two thermal models  is motivated by the expectation of an accretion disk that reaches all the way onto the NS surface on which it forms a layer of hot optically thick plasma. This assumption relies on long established theoretical considerations \citep[e.g.,][]{1986SvAL...12..117S,2000AstL...26..699S} and a strong observational record of the presence of two thermal components in soft state NS-XRBs with low magnetic fields \citep[e.g.,][]{1984PASJ...36..741M,1988ApJ...324..363W,1989PASJ...41...97M,2001AdSpR..28..307B,2007ApJ...667.1073L,2013MNRAS.434.2355R}.

Despite the simplicity of our spectral models, we are confident that their parameters describe actual physical characteristics of the system (i.e.~temperature and size of emitting regions). Therefore we pay  particular attention to the accuracy of fit derived values. We are particularly interested in the ionization state of the optically thick matter, while ensuring that the derived quantities are sensible in the context of accreting low magnetic field (low-B) NSs, i.e., temperature and inner disk radius commensurate with the size of the NS and  the formation of the layer of hot (${\gtrsim}$1\,keV) optically thick plasma on its surface \citep[e.g.,][]{2001ApJ...547..355P}.  For these reasons it is important to take into account the effects of very high temperature (${\gtrsim}10^6$\,K) on the spectral shape of the thermal emission.

At this temperature range, most of the lighter elements in the upper disk layers will become fully ionized (see Section~\ref{sec:discussion}) resulting in an abundance of hot free electrons that significantly increase the impact of photon-electron scattering within the disk and boundary layer. The process modifies the black body spectrum, which -- to the first approximation -- appears "harder" than it actually is. With respect to the spectral analysis -- when modelling the continuum with typical thermal models such as  {\tt diskbb} or {\tt bbodyrad}  -- our estimations of disk temperature and radius will deviate from their actual values. This issue is often called "spectral hardening" and is usually addressed by applying a correction factor ($f_{\rm col}$) to  the temperature and normalization of the thermal models (i.e.~{\tt diksbb} and {\tt bbodyrad} \citep{1986ApJ...306..170L,1986SvAL...12..383L,1995ApJ...445..780S}. Namely,
\begin{equation}
T_{\rm cor}=\frac{T_{\rm in}}{f_{\rm col}}
\label{eq:Tcor}
\end{equation}
and
\begin{equation}
R_{\rm cor}=R_{\rm in}\,{f_{\rm col}}^2
\label{eq:Rcor}
\end{equation}
where $T$ is the temperature of the MCD component and $R_{\rm in}$ is the inner radius. The value of $f_{\rm col}$ has been obsessionally constrained between $\sim1.5$ and $\sim$2.1 \citep[e.g.,][and references therein]{2005ApJ...618..832Z}.  In the following analysis we present all best fit values without correcting for the expected spectral hardening (i.e.~$f_{\rm col}{=}1.0$).the effects of a more realistic value of $f_{\rm col}{\sim}1.8$ on the best-fit estimated values of disk temperature and inner radius are discussed in Section~\ref{sec:discussion}.}

\subsection{X-ray spectroscopy of 4U 1543-624}

\subsubsection{{\xmm} April 2001 observation}
\label{sec:XMM}

\begin{figure}
\includegraphics[angle=-90, width=\columnwidth]{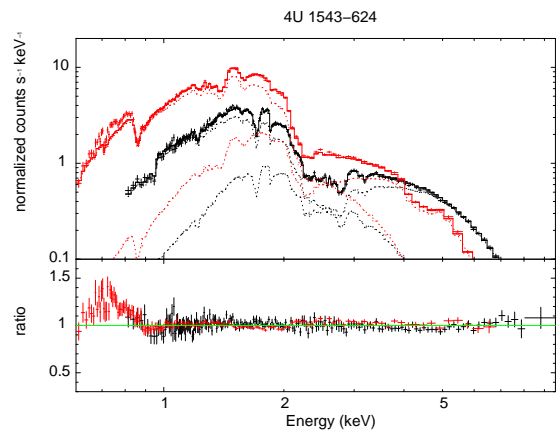}
\caption[]{Normalized counts vs. energy and ratio of the data to the continuum  model for the 2000 Chandra observation. The data have been re-binned for clarity; 
the 1-10\,keV energy range is shown. }
\label{fig:Chandra}
\end{figure}
 
\begin{figure}
\includegraphics[angle=-90, width=\columnwidth]{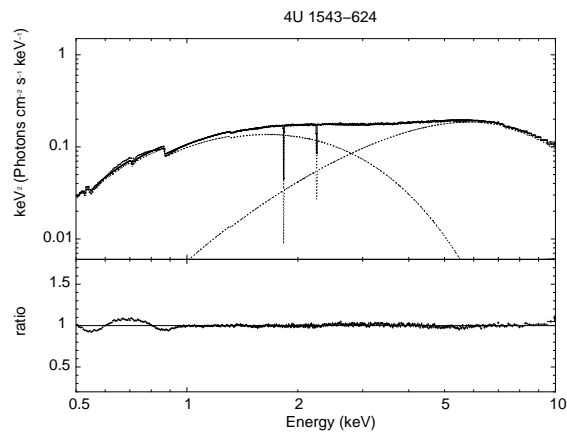}
\caption[Ratio of the data to the continuum model  for {\src}]{Unfolded spectrum vs. energy and ratio of the data to the continuum  model for the 2001 {\xmm} observation. { The data have been re-binned for clarity; the full 0.5-10\,keV energy range is shown. The unfolded spectrum was chosen in order to visually demonstrate that the soft thermal component can be well constrained even if we ignore energy channels below 1\,keV} }
\label{fig:XMM_1543_ld}
\end{figure}

\begin{figure}
 \includegraphics[angle=-90, width=\columnwidth]{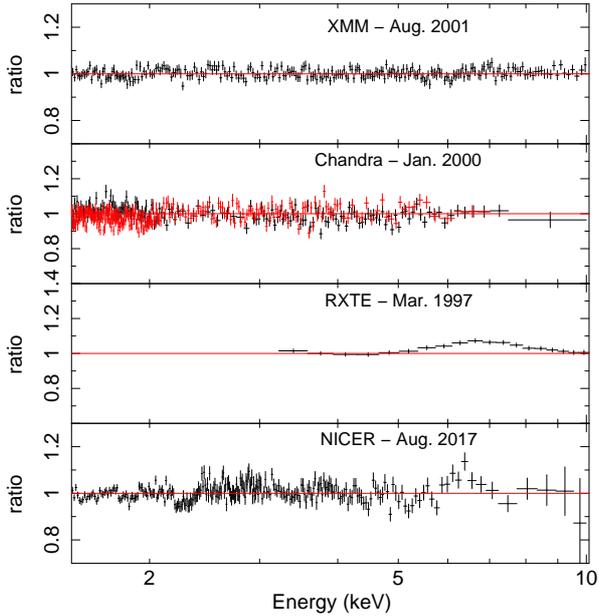}
 \caption{ Data-to-model ratio plots for {\src} illustrating the Fe K${\alpha}$ emission line variability through different epochs.}
 \label{fig:variable_1543}
\end{figure}

{\xmm} observed {\src} in April of 2001. All onboard instruments were operational. Namely, MOS1 and pn were operating in Timing Mode and MOS2 in Small Window Mode. All detectors had the Medium optical blocking filter on. In Timing mode, data are registered in one dimension, along the column axis, which results in significantly shorter CCD readout time (0.06\,ms for Timing mode, instead of 5.7\,ms of the Small Window mode). In addition to increased time resolution, the use of Timing mode may protect observations of bright sources from pile-up\footnote{For more information on pile-up see: http://xmm2.esac.esa.int/docs/documents/CAL-TN-0050-1-0.ps.gz}. 
\\
\\
{\it Spectral extraction and analysis.}\\
Spectra from all detectors were extracted using the latest {\xmm} Data Analysis software SAS, version 15.0.0. and using the calibration files released\footnote{XMM-Newton CCF Release Note: XMM-CCF-REL-334}  on May 12, 2016. All observations were checked for high background flaring activity. To this end, we extracted high energy light curves (E$>$10\,keV for MOS and 10$<$E$<$12\,keV for pn) with a 100\,s bin size. A review of the lightcurves revealed no evidence of high energy flares in any of the detectors. The spectral extraction was done using SAS task \texttt{evselect}, with the standard filtering flags (\texttt{\#XMMEA\_EP \&\& PATTERN<=4} for pn and \texttt{\#XMMEA\_EM \&\& PATTERN<=12} for MOS). SAS tasks \texttt{rmfgen} and \texttt{arfgen} were used to create the redistribution matrix and ancillary file. MOS2 data suffered from pile-up and for this reason they were rejected. Furthermore, because the effective area of pn at $\sim$7\,keV is approximately five times higher than that of MOS and since the main interest of this work lies in this energy range, we opted to only use the pn data for our analysis. \src is known to exhibit a complex emission and an absorption-like spectrum below 1\,keV.

Several studies have indicated the presence of strong emission lines below 1\,keV, as well as pronounced absorption edges, which are attributed to fluorescent emission and/or absorption from highly non-solar C/O or O/Ne/Mg-rich material transferred onto the NS by its degenerate donor \citep[e.g.,][]{2001ApJ...560L..59J,2003ApJ...599..498J,2011MNRAS.412L..11M}. The spectral models described in this section were applied to the entire energy range available from all detectors and we confirm the presence of soft emission-like features consistent with highly ionized oxygen. The emission-like feature is presented in Figure~\ref{fig:Chandra} for the \cxo observation of \src (i.e.,~normalized counts and data-to-model ratio vs energy, without accounting for the ${\sim}$0.68keV emission line reported by previous authors) and in Figure~\ref{fig:XMM_1543_ld} for the \xmm data { where again the broad emission-like feature and potential absorption edges are evident in the ratio plot. Modeling the residuals with a gaussian emission line confirms the presence of a line centered at ${0.68{\pm0.01}}$\,keV with a width of 70.9${\pm}0.25$\,eV (in the 90\% confidence range).  Nevertheless, the analysis of these features is not the focus of our study and since ignoring the energy channels below 1\,keV simplifies our analysis without affecting our parameter estimations, all tabulated best-fit values are for the spectral analysis of energy channels above 1\,keV. This decision was made after confirming that the best-fit values for the soft thermal component (\texttt{diskbb} model) are well constrained in the 1-10\,keV range. Indeed our broadband analysis yielded kT$_{\rm in}$ and disk normalization values consistent, within the 90\% confidence range, with the 1-10\,keV fit. The best-fit values are also consistent within broadband models that include or omit the 0.68\,keV gaussian emission line.} For a detailed study of the atomic features and absorption edges in the low energy part of the spectrum of \src, we refer the reader to the works of \cite{2003ApJ...599..498J}, \cite{2011MNRAS.412L..11M} and most recently \cite{2019ApJ...883...39L}.

We fit the 1--10\,keV spectral continuum with a combination of absorbed black body (xspec model \texttt{bbodyrad}) and disk black body (xspec model \texttt{diskbb}) components. The interstellar absorption was modeled using the improved version of the \texttt{tbabs} code\footnote{http://pulsar.sternwarte.uni-erlangen.de/wilms/research/tbabs/} \citep{2000ApJ...542..914W}. The hydrogen column density (nH) was frozen at the value provided by the HEASARC nH tool \citep{2016A&A...594A.116H}.  The fit yields a reduced $\chi^2$ value of 1.04 for 1858/1795 dof, and the data-to-model ratio plot does not indicate emission-like  residuals in the entire 1--10\,keV region (Figure~\ref{fig:variable_1543}). Temperature of the disk black body is $kT_{\rm in}{\sim}$0.65 \,keV and the black body temperature is $kT_{\rm BB}{\sim}$1.53\,keV. There is no evidence of iron K$\alpha$ emission, with a $1\,{\sigma}$ upper limit of 2.68\,eV on the EW of the line. The source luminosity, extrapolated in the 0.50-30\,keV range, is $\approx5.1\times10^{36}$\,erg/s, for a distance of 7\,kpc \citep{2004ApJ...616L.139W}. All values of the best fit parameters are presented in Table~\ref{tab:cont1543}.

\subsubsection{RXTE May 1997 observation}
\label{RXTE}

\begin{figure}
\includegraphics[angle=-90, width=\columnwidth]{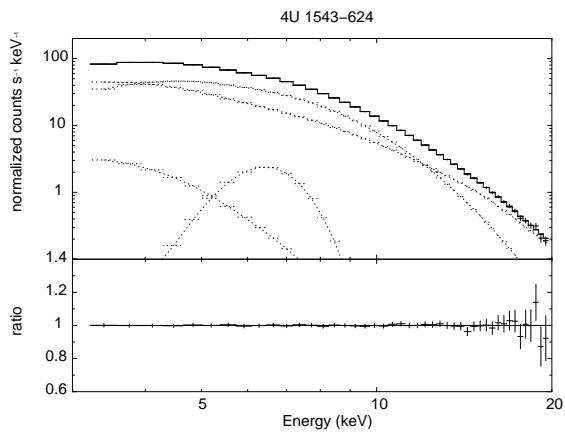}
\caption[Ratio of the data to the continuum model  for {\src}]{Normalized counts vs. energy and ratio of the data to the continuum model for the 1997 RXTE observation. The data have been re-binned for clarity; 
the 3-20\,keV energy range is shown. }
\label{fig:RXTE_1543_ld}
\end{figure}

There are two RXTE observations of {\src}, both taken in 1997. The first observation commenced on May 5th and the second on the 22nd of September. Both observations were broken up into multiple intervals. We have analyzed all available observations. However, for better clarity and brevity, we only present the results of the longest (14\,ks) observation, taken on May 6th 1997 (P20071-04-01-00). 
\\
\\
{\it Spectral extraction and analysis.}\\ 
RXTE had three on board instruments: the All Sky Monitor  \citep{1996ApJ...469L..33L}, the Proportional Counter Array \citep[PCA,][]{2006ApJS..163..401J} and the High Energy X-ray Timing Experiment \citep[HEXTE,][]{1998ApJ...496..538R}. 
In this work we obtain and analyze the source spectrum from the PCA, which provides the highest energy resolution (18\% at 6\,keV) in the spectral range of interest (3--20\,keV). The PCA detector is comprised of five Proportional Counter Units (PCUs), with a combined effective area of $\sim7000\,{\rm cm}^2$. Each PCU is composed of three layers of xenon (90\%) and methane (10\%) composites. Most incident photons with energies below 20\,keV are detected in the top layer (layer 1) and for this reason we have opted to only extract spectra from this layer. All PCUs were active during the observation, which allowed us to obtain the data of all detectors for the spectral extraction. As per the recommendation\footnote{https://heasarc.gsfc.nasa.gov/docs/xte/pca/doc/rmf/pcarmf-11.7/} of the RXTE guest observer facility (GOF), we removed energy channels below 3\,keV. Additionally, in our spectral extraction we ignored all channels above 20\,keV. The signal-to-noise ratio drops significantly above this threshold, and for this work we are primarily interested in the energy range below 20\,keV. Due to the low orbit of RXTE, observations are occasionally affected by the satellite's passage through the South Atlantic Anomaly (SAA) or by Earth occultation of the observed source. Data may also be affected by electron contamination and sporadic breakdowns of the PCU2 detector. For the spectral extraction we filtered out intervals where the elevation angle was less than 10 degrees (to avoid possible Earth occultations) and also any data that may have been received during passage from the SAA. Furthermore, we excluded any data affected by electron contamination, where offset by more than 0.02 degrees or taken 150 seconds before, through 600 seconds after a PCU2 breakdown. Source spectra were extracted and background emission was modeled using standard routines from the {\small FTOOLS} package and the latest model for bright sources, provided by the GOF\footnote{http://heasarc.gsfc.nasa.gov/docs/xte/pca\_news.html}. 
All spectra were re-binned to a minimum of 30 counts per bin. 

The spectral continuum is again modeled with a similar combination of absorbed disk black body and black body emission, and a hydrogen column density (nH) frozen at the Galactic value. The fit yields a reduced $\chi^2$ value of 1.66 for 41 dof, and the data-to-model ratio plot reveals strong positive residuals in the 6-7\,keV region (Figure~\ref{fig:variable_both}). The residual structure -- a strong indication of a bright iron K$\alpha$ emission line --  is modeled using a Gaussian. The high energy range of RXTE/PCA ($>10$\,keV) allowed us to also probe the non-thermal tail of the \src spectrum, which we modeled using a power law with a spectral index of ${\sim}2.5$ and a high energy exponential cutoff at ${\sim}15$\,keV. The final fit yields a reduced $\chi^2$ value of 0.44 for 38 dof (Figure \ref{fig:RXTE_1543_ld} for the 3-20\,keV best-fit plot). The  temperature of the disk black body is $kT_{\rm in}{\sim}0.8$\,keV and for the black body is $kT_{\rm BB}{\sim}1.8$\,keV. The iron emission line is centered at  ${\sim}6.58$\,keV, has a width of ${\sim}$670\,eV and an equivalent width (EW) of ${{\sim}}100$\,eV. The source luminosity, extrapolated to the 0.50-30\,keV range, is $\approx7.1\times10^{36}$\,erg/s. All values of the best fit parameters are presented in Table~\ref{tab:cont1543}. Findings from the spectral analysis  of the \xmm observation (subsection \ref{sec:XMM}) and RXTE observation (subsection \ref{RXTE}) are also briefly presented in our comprehensive analysis of the entire UCXB catalog in \cite[][accepted for publication in MNRAS]{2020arXiv200100716K}.

\subsubsection{Chandra December 2000 observation}

The \cxo X-ray Observatory observed {\src} four times. Once in December of 2000, using the HETG and three more times in June 2012, using the Low Energy Transmission Grating (LETG). Below we outline the spectral extraction and analysis for the 2000 HETG observation. 
\\
\\
{\it Spectral extraction and analysis.}\\
Using the tools included in the latest version of the CIAO software (vers. 4.8), we extract spectra from both the medium energy grating (MEG) and the high energy grating (HEG). Namely, upon selecting an appropriate extraction region and  taking into account the correct source position and telescope orientation during the entire observation, we use CIAO task \texttt{tgextract}  to create a standard type II spectral file, containing all spectral orders of both the MEG and HEG detectors, and \texttt{mktgresp} to create their corresponding ancillary files and response matrices. In order to analyze the spectra with XSPEC, we create separate type I PHA files for the MEG and HEG spectra, simultaneously adding the +1 and -1 grating orders and rebinning them to a minimum of 30 counts per bin. We also create a separate background spectrum using the task \texttt{tg\_bkg}.

The spectral continuum is well modeled with the same combination of absorbed disk black body and black body emission, as in the 1997 RXTE and 2001 {\xmm} observations. The temperature of the disk black body lies at $kT_{\rm in}{\sim}$0.65\,keV and the black body temperature at $kT_{\rm BB}{\sim}$1.61\,keV. This time, no emission line was detected in the 6-7\,keV range and we place a $1\,{\sigma}$ upper limit for the EW of the iron K${\alpha}$ emission line at 6.13\,eV. The double thermal model fit yielded a reduced $\chi^2$ value of 1.02 for 3703 dof. The source luminosity during the Chandra observation, was $\approx6.1\times10^{36}$\,erg/s, calculated in the 0.50-30\,keV range and assuming a distance of 7\,kpc. All values of the best fit parameters are presented in Table~\ref{tab:cont1543}.

\begin{table}
\centering
 \caption{List of observations of \src and \srcs analyzed in  this paper}
 \label{tab:obs2}
 \begin{tabular}{@{}lcccc}
 Instrument & obsID & Date &  Duration$^{1}$ \\ 
            &       &      &   (ks) \\
 \hline
  \hline
{\src}\\
 \hline
   {\it RXTE}       & P20071    &1997-05-06  & 14\\  
   {\it Chandra}    & 702       &2000-09-12  & 30 &    \\   
   {\it XMM-Newton} & 0061140201&  2001-02-04  & 50\\%
   {\it NICER     } & 1050060106&  2017-08-20  & 13\\%
  \hline
{\srcs}\\
 \hline
   {\it RXTE}       & P92050     & 2007-06-16  & 12\\%
   {\it XMM-Newton} & 0830190401 & 2018-04-08  & 65 &    \\ 
   {\it {\nus}}     & 90402313002 & 2018-04-08 & 40 &    \\ 
  \hline
\end{tabular}

 \medskip
{$^{1}$Duration of filtered observations.\\}
\end{table}

\subsubsection{NICER Sept. 2017 observation}

 \src was observed with NICER in 13 separate pointings over $\sim$10 days starting on August 15 2017. Here we analyse obsID 1050060106, which at a duration of ${\sim}$13\,ks, provides the highest number of total counts after the standard event filtering described below.
\\
\\
{\it Spectral extraction and analysis.}\\
The {\it Neutron Star Interior Composition Explorer} (NICER; \citealt{2012SPIE.8443E..13G}) operates on-board the International Space Station and consists of 56 ``concentrator" optics and silicon drift detector pairs registering X-ray photons in the $0.2-12$\,keV energy range. The 52 operating collectors comprise a total collecting area of ${\approx}1900$\,cm$^{2}$ at 1.5\,keV. {\sc nicerdas} 2018-10-07\_V005 was employed for the data reduction process. Namely, using  {\sc nimaketime}  we created appropriate good time intervals (GTIs), by filtering out all raw data with the standard criteria using various housekeeping parameters (see \citealt{bogdanov19,guillot19}, for more details). The final event list was extracted using {\sc niextract-events} for PI energy channels between 40--1200, inclusive, and EVENT\_FLAGS=bxxx1x000 as per the NICER manual guidelines. For the spectral extraction we used the heasoft tool {\sc xselect}. The spectra were subsequently normalized based on the instrumental residuals calibrated from the Crab Nebula, as described in \cite{ludlam18}. 

The background spectrum was generated from a library of \nicer\ observations of "blank sky" fields (the same ones as for RXTE, \citealt{jahoda06}). Specifically, a weighted-average of these "blank sky" observations with a similar combinations of observing conditions as our target's data set is used to produce the diffuse cosmic X-ray background spectrum\footnote{This background modeling technique is detailed in \cite{bogdanov19}}. The latter is also normalized with the Crab Nebula residual spectrum. Finally, the latest ARF and RMF files were kindly provided by the NICER GOF (these are now publicly available as NICER XTI Calibration Files: 20200722 in the heasarch website).

Observed by NICER, almost two decades after the {\xmm} observations, the spectral continuum of {\src} is  qualitatively the same as in all previous observations and was again modeled with the same combination of absorbed disk black body and black body emission. The accretion disk temperature of the disk black body is closer to its value during the RXTE observation, at $kT_{\rm in}{\sim}$0.80\,keV as is the black body temperature which is estimated at $kT_{\rm BB}{\sim}$1.80\,keV. The iron  K${\alpha}$ emission line has reappeared in the source spectrum, indicated by positive residual structure in the data-to-ratio vs energy plot in Figure~\ref{fig:variable_1543}. The emission line was modeled with a Gaussian centered at ${\sim}6.5$\,keV, with a width of ${\sim}$160\,keV  and an EW of ${\sim}61$\,eV. The double thermal model including the Gaussian emission line yielded a reduced $\chi^2$ value of 1.07 for 743 dof. The source luminosity during the NICER observation, was $\approx6.0\times10^{36}$\,erg/s, calculated in the 0.50-30\,keV range and assuming the distance of 7\,kpc. All values of the best fit parameters are presented in Table~\ref{tab:cont1543}.

\begin{table*}
 \centering
 \begin{minipage}{128mm}
\caption[Best fit parameters]{Best fit parameters for the {\xmm}, Chandra, RXTE/PCA and NICER spectra of {\src}. The errors are in the  90\% confidence range.}
\label{tab:cont1543}
  \begin{tabular}  {lccccccc}
  \hline
Model parameter                  & RXTE-PCA 1997$^{\it a}$  & Chandra 2000              &  {\xmm} 2001          & NICER 2017  \\ 
 \cline{1-5}                                                                                                                   
 \hline                                                                                                                    
nH$^{\it b}$ ($10^{22}\,{\rm cm^{-2}}$) & 0.29              & 0.29                      &  0.29                      &  0.29   \\
\hline                                                                                                                     
                                                                                                                           
{\em Disk Black Body}    \\                                                                                                  
  \hline                                                                                                                   
 kT$_{\rm in}$ (keV)                       & 0.80$_{-0.22}^{+0.31}$    & 0.71$_{-0.003}^{+0.004}$  &  0.65$\pm0.02$             & 0.79$_{-0.03}^{+0.04}$   \\
 ${R_{\rm in}}^{\it d}$   (km)  & 9.02${\pm}{1.71}$    & 12.2$_{-0.26}^{+0.30}$     &  13.6$_{-0.66}^{+0.74}$    & 10.3$_{-0.42}^{+0.54}$    \\
 \hline                                                                       
{\em Black Body}   \\                                                                                                        
  \hline                                                                                                                   
  kT$_{\rm BB}$ (keV)   			& 1.81$_{-0.13}^{+0.24}$    & 1.61$_{-0.03}^{+0.04}  $  & 1.53$_{-0.006}^{+0.005} $  & 1.80${\pm}0.02$    \\
  norm$^{\it e}$                & 7.48$_{-0.98}^{+0.27}$    & 7.00${\pm}0.10         $  & 9.54$_{-0.76}^{+0.79}$     & 4.26${\pm}0.15$    \\
  \hline       
{\em Iron Line}    \\                                                                                                        
  \hline                                                                                                                   
  Centroid E (keV)              & 6.58$_{-0.33}^{+0.21}$    & 6.6$^{f}$             & 6.6$^{f}  $            & 6.47${\pm}0.12$       \\
  Width $\sigma$(eV)            & 678$_{-121}^{+131}$       & $500^{f}$                 &   $500^{f}$                & 160$_{-7.05}^{+18.2}$ \\
  Flux$^{\it g}$                & 95.1$_{-23.5}^{+34.5}$    & $<5.05$                   &     $<1.19$                & 31.5${\pm}5.7$\\
  EW (eV)  		             	& 99.3$_{-24.3}^{+36.6}$    & $<6.13$                   &     $<2.68$                & 61.3${\pm}11.7$     \\ 
\hline                                                                                                                     
  ${L_{Total}}^{\it h}$         & 6.65${\pm}{0.03}$         & 6.14$_{-0.08}^{+0.09}$    &  5.14${\pm}0.01$            &  6.00${\pm}0.01$             \\
  ${L_{DBB}}^{\it h}$           & 1.56$_{-0.66}^{+0.46}$    & 2.48$_{-0.10}^{+0.32}$    &  2.78${\pm}0.01$            &  3.35$_{-0.12}^{+0.18}$             \\
  ${L_{BB}}^{\it h}$            & 3.82$_{-0.44}^{+0.64}$    & 3.76${\pm}{0.03}$        &  2.37${\pm}0.01$            &  2.54${\pm}{0.09}$           \\
  ${L_{BPL}}^{\it h}$           & 1.13${\pm}{0.14}$         & --                       & --                          &   --         \\
                                                                                                                           
\hline                                                                                                                     
$\chi^2/{\rm dof}$              & 16.9/38                   & 3762/3703                 &  1858/1795                  & 807/743  \\
\end{tabular}
 \medskip

{ $^{\it a}$ With an additional cutoff power law (${\Gamma}{=}2.49_{-0.08}^{+0.17}$ and $E_{cut}{=}14.8_{-0.85}^{+0.23}$\,keV), for the RXTE data \\
  $^{\it b}$ Parameter frozen at total galactic H\,I column density provided by the HEASARC nH tool \citep{2016A&A...594A.116H}.\\
 $^{\it c}$ $10^{-3}\,{\rm ph\,keV^{-1}\,cm^{-2}\,s^{-1}}$.\\
 $^{\it d}$ Solving K=${\rm (1/{f_{\rm col}}^{4})((R_{in}/D)^{2}_{10kpc}})\,\cos{i}$, for $R_{in}$ (the inner radius of the disk in km).  K is the normalization of the \texttt{diskbb} model, $f_{\rm col}$ is the spectral hardening factor, ${\rm D_{10kpc}}$ is distance in units of 10\,kpc and $i$ is the inclination. In the tabulated values we have set $f_{\rm col}=1$.  \\
 $^{\it e}$  ${\rm(R_{\rm bbody}/{D_{10kpc}})^{2}}$ where ${\rm R_{\rm bbody}}$ is the size of the emitting region in km and ${\rm D_{10kpc}}$ is distance in units of 10\,kpc. \\
 $^{\it f}$ Parameters frozen. The centroid energy value was frozen at the median value of the 6.4-6.9\,keV range, for a step of 0.1\,keV  and width value based on average width of observed Fe K$\alpha$ emission lines in LMXBs \citep[e.g.,][]{2009ApJ...690.1847C,2010A&A...522A..96N}.\\
 $^{\it g}$  ${\rm 10^{-5}\,ph\,cm^{-2}s^{-1}}$.\\
 $^{\it h}$ Luminosity in units of $10^{36}$\,erg/sec, extrapolated to the 0.50-30\,keV range and assuming a distance of 7\,kpc \citep{2003A&A...397..249S,2004ApJ...616L.139W}. \\}

\end{minipage}
\end{table*}

\subsection{X-ray spectroscopy of Swift J1756.9-2508}

\subsubsection{RXTE June 2007 observation}

\begin{figure}
\includegraphics[angle=-90, width=\columnwidth]{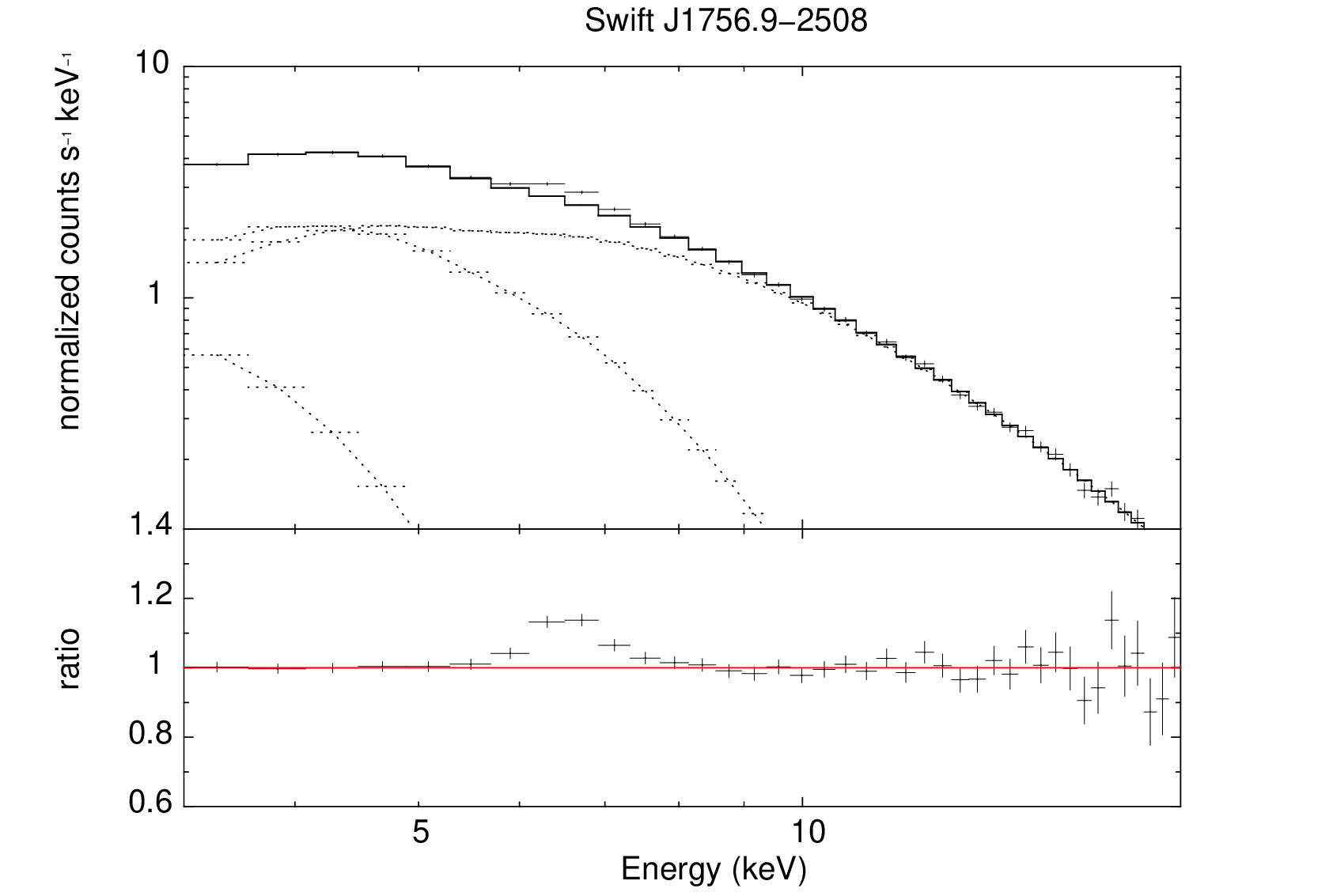}
\caption[]{Normalized counts vs. energy and ratio of the data to the continuum  model for the 2007 RXTE observation {\srcs}. The data have been re-binned for clarity; 
the 1-10\,keV energy range is shown. }
\label{fig:RXTE_ld_Swift}
\end{figure}

\begin{figure}
 \includegraphics[angle=-90, width=\columnwidth]{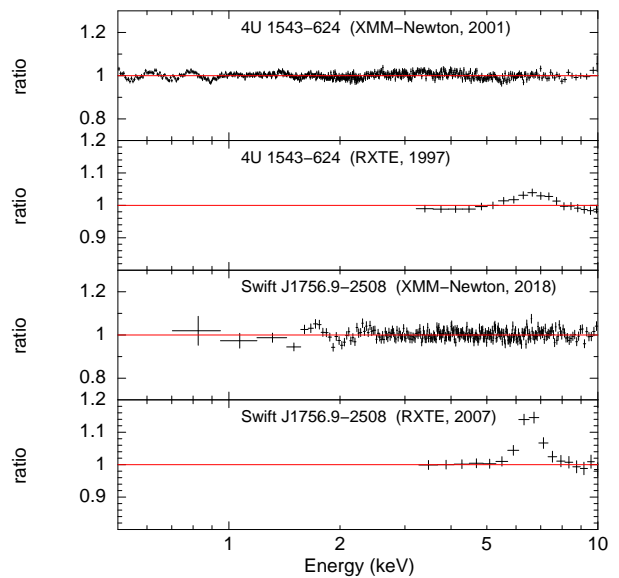}
 \caption{ Data-to-model ratio plots for the two sources with a variable Fe K${\alpha}$ emission line.}
 \label{fig:variable_both}
\end{figure}

There are two sets of RXTE observations of \srcs carried out in 2007 when the source was first detected, and in 2009 during its second outburst. We have reviewed all available data and present here our analysis of the 2007 observation with the higher number of registered counts. For a thorough study of the entire RXTE observation campaign of \srcs we refer the reader to the works of \cite{2007ApJ...668L.147K} and \cite{2010MNRAS.403.1426P}.
\\
\\
{\it Spectral extraction and analysis.}\\
The extraction procedure for the RXTE/PCA spectra of \srcs is identical to the one described in paragraph~\ref{RXTE}.
The X-ray spectrum of \srcs during its 2007 outburst is also described by a combination of two absorbed thermal components, which is consistent with the high-luminosity soft-state spectra of weakly magnetized NS LMXBs \citep[e.g.,][]{2007ApJ...667.1073L}. As expected for a source in the direction of the Galactic bulge there is considerable interstellar absorption in the line of sight to \srcs, with nH${\sim}6{\times}10^{22}$\,cm$^{s}$ \citep{2016A&A...594A.116H}. The accretion disk temperature was estimated at $kT_{in}{\sim}$1.1\,keV and the temperature of the black body component at $kT_{BB}{\sim}$2.3\,keV. 
A prominent iron  K${\alpha}$ emission line is clearly detected above the spectral continuum (see Figure~\ref{fig:variable_both} for the data-to-model vs energy plot). The emission line is centered at ${\sim}6.5$\,keV, and has an EW of ${\sim}171$\,keV. The width of the emission line is smaller than the RXTE/PCA energy resolution (18\% at 6\,keV) and was thus frozen to 500\,eV. In addition to the thermal emission, the fit required a non-thermal component dominating ${>}10$\,keV, which was modelled using a  power law with spectral index of ${\Gamma}{\sim}1$ and a high-energy exponential cutoff at ${\sim}12$\,keV. The model yields a reduced $\chi^2$ value of 0.65 for 30 dof. The source luminosity during the RXTE observation, was $\approx5\times10^{36}$\,erg/s, calculated in the 0.50--30\,keV range and assuming a distance of 8.5\,kpc,  based  on  the  source position toward  the direction of the Galactic centre. All values of the best fit parameters along with errors in the 90\% confidence range are presented in Table~\ref{tab:contswift}.

\begin{table}
 \caption[Best fit parameters]{Best fit parameters for the combined {\nus} -- {\xmm} and RXTE/PCA spectra of {\srcs}. The errors are 90\%.}
\label{tab:contswift}
 \begin{center}
 \begin{tabular}  {lccccc}
  \hline
Model parameter                           &  RXTE/PCA                & XMM-{\nus}    \\  
                                          &  2007                     &  2018  \\ 
 \cline{1-3}                                                            
 \hline                                                             
nH$^{\it a}$ ($10^{22}\,{\rm cm^{-2}}$)   &  6.05$\pm1.02$            &  7.28$\pm0.07$   \\
\hline                                                              
                                                                    
{\em Disk Black Body}    \\                                           
  \hline                                                            
 kT$_{\rm in}$ (keV)                                 &  0.96$_{-0.18}^{+0.39}$    & 1.24$_{-0.06}^{+0.09}$   \\
 ${R_{\rm in}}^{\it b}$   (km)            &  3.9$_{-1.66}^{+1.04}$    & 5.45$_{-0.42}^{+0.45}$    \\
 \hline                                  
{\em Black Body}   \\                                                          
  \hline                                                                     
  kT$_{\rm BB}$ (keV)   			                  & 2.27$_{-0.09}^{+0.14} $    & 2.79$_{-0.12}^{+0.17}$    \\
  norm$^{\it c}$                          & 12.3$_{-6.53}^{+3.46}$     & 1.12${\pm}0.15$    \\
  \hline  
  {\em Cutoff Power Law}                                                                      \\
  \hline
  $\Gamma$   		                      & 0.97$_{-0.25}^{+0.22} $   & 1.39$_{-0.09}^{+0.07} $ \\
  norm (${\times}10^{-3}$)                & 11.6$_{-0.77}^{+1.17} $   & 20.1$_{-4.89}^{+5.11}$   \\
  $E_{cut}$  (keV)  		                  & 11.9$_{-0.69}^{+0.71} $   & 22.8$_{-0.68}^{+1.01}$      \\
  $E_{fold}$ (keV)                        & 17.4$_{-2.67}^{+3.41} $   & 50.6$_{-6.38}^{+8.27} $       \\
 \hline  
{\em Iron Line}    \\                                                          
  \hline                                                                     
  Centroid E (keV)                        & 6.47${\pm}0.12$           & 6.6$^{d}  $   \\
  Width $\sigma$(eV)                      &   $500^{d}$               & $500^{d}$  \\
  Flux$^{\it e}$                          &  42.3${\pm}4.5$           & $<1.19$\\
  EW (eV)  		                          &  171${\pm}18.5$           &   $<7.36$  \\ 
\hline                                                                       
${L_{Total}}^{\it f}$ (${\times}10^{36}$\,erg/s)  & 4.87${\pm}{0.10}$         & 3.67${\pm}{0.02}$  \\
${L_{DBB}}^{\it f}$ (${\times}10^{36}$\,erg/s) & 0.84${\pm}{0.11}$         & 0.91${\pm}{0.07}$  \\
${L_{BB}}^{\it f}$  (${\times}10^{36}$\,erg/s) & 1.00${\pm}{0.12}$         & 0.67${\pm}{0.04}$  \\
${L_{cPL}}^{\it f}$ (${\times}10^{36}$\,erg/s) & 3.13${\pm}{0.12}$         & 2.59${\pm}{0.25}$  \\
\hline                                                              
$\chi^2/{\rm dof}$                        &  19.6/30                  & 3255/3141  \\
\end{tabular}                    
 \medskip
 \end{center}
{ $^{\it a}$ $10^{-3}\,{\rm ph\,keV^{-1}\,cm^{-2}\,s^{-1}}$.\\
 $^{\it b}$ Solving K=${\rm (1/{f_{\rm col}}^{4})((R_{in}/D)^{2}_{10kpc}})\,\cos{i}$, for $R_{in}$ (the inner radius of the disk in km).  K is the normalization of the \texttt{diskbb} model, $f_{\rm col}$ is the spectral hardening factor (again for the tabulate values $f_{\rm col}=1$), ${\rm D_{10kpc}}$ is distance in units of 10\,kpc and $i$ is the inclination. \\
 $^{\it c}$  ${\rm(R_{\rm bbody}/{D_{10kpc}})^{2}}$ where ${\rm R_{\rm bbody}}$ is the size of the emitting region in km and ${\rm D_{10kpc}}$ is distance in units of 10\,kpc. \\
 $^{\it d}$ Parameters frozen. The centroid energy value was frozen at the median value of the 6.4-6.9\,keV range, for a step of 0.1\,keV  and width value based on average width of observed Fe K$\alpha$ emission lines in LMXBs \citep[e.g.,][]{2009ApJ...690.1847C,2010A&A...522A..96N}, for the RXTE detection the width is frozen at the instrument's energy resolution at 6\,keV. The two values coincide.\\
 $^{\it e}$  ${\rm 10^{-5}\,ph\,cm^{-2}s^{-1}}$.\\
 $^{\it f}$ Luminosity extrapolated to the 0.50-30\,keV range and assuming a distance of 8.5\,kpc, based on the proximity toward the direction of the Galactic centre . \\}

 \end{table}

\begin{figure}
 \includegraphics[angle=-90, width=\columnwidth]{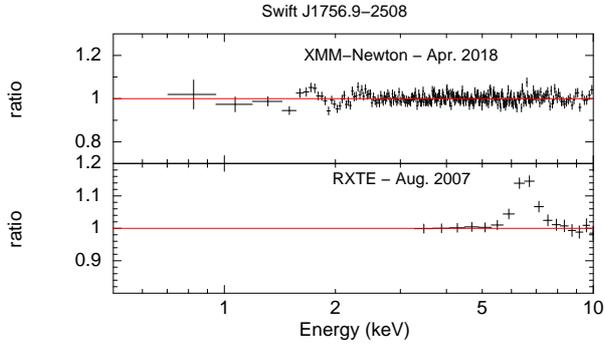}
 \caption{ Data-to-model ratio plots for {\srcs} illustrating the Fe K${\alpha}$ emission line variability through different epochs.}
 \label{fig:variable_swift}
\end{figure}

\subsubsection{{\it XMM-Newton/NuSTAR} April 2018 observation}

\begin{figure}
\includegraphics[angle=-90, width=\columnwidth]{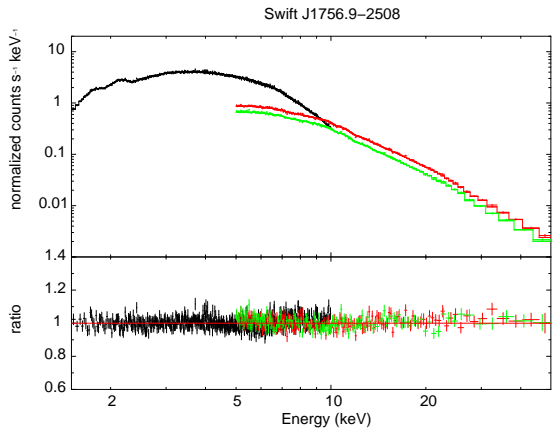}
\caption[]{Normalized counts vs. energy and ratio of the data to the continuum  model for the 2018 {\it XMM-Newton/NuSTAR}  observation {\srcs}. The data have been re-binned for clarity; 
the 1-50\,keV energy range is shown. }
\label{fig:swift_ld_noadd}
\end{figure}

\srcs was jointly observed with {\xmm} and {\nus} X-ray telescopes during its third and latest outburst in April of 2018.  
\\
\\
{\it Spectral extraction and analysis.}\\
During the {\xmm} observation of {\srcs} the EPIC-pn detector was again operated in Timing mode and therefore the spectral extraction process is the same as in \src (paragraph \ref{sec:XMM}). For the \nus\ data extraction, we used version 1.9.3 of the \nus\ data analysis system (\nus\ DAS) and the latest instrumental calibration files from CalDB v20191008. Data were cleaned and calibrated using the \texttt{NUPIPELINE} routine with default settings. The internal high-energy background was reduced and  passages through the South Atlantic Anomaly were screened (settings SAACALC$=$3, TENTACLE$=$NO and SAAMODE$=$OPTIMIZED). Phase-averaged source and background spectra were extracted using the \texttt{NUPRODUCTS} script, which also produces the instrumental responses for both focal plane modules, FPMA and B. We used a circular extraction region with an 80{\arcsec}  radius for both the source and the background spectra. The latter were extracted from a blank sky region in the same detector as the source and at an adequate distance from it in order to avoid any contribution from the PSF wings. The default PSF, alignment, and vignetting corrections were used. 

During the joint {\it XMM-Newton/NuSTAR} observation of 2018, \srcs was again detected in the soft-state, with a qualitatively identical spectrum to the 2007 (this work) and 2009 \citep{2010MNRAS.403.1426P} RXTE observations.  Exploiting the energy range of \xmm we update the estimation of interstellar absorption to a value of nH${\sim}7{\times}10^{22}$\,cm$^{-2}$. The accretion disk temperature was estimated at $kT_{in}{\sim}$1.2\,keV and the temperature of the black body component at $kT_{BB}{\sim}$2.8\,keV.
The non-thermal, power-law shaped component of the spectrum had a spectral index of ${\Gamma}{\sim}1.4$ and an exponential drop-off at ${\sim}23$\,keV. Despite the similarity in the spectral continuum, the prominent Fe K${\alpha}$  emission line detected in both the 2007 and 2009 RXTE observations had disappeared in the 2018 observation. We place a $1\,{\sigma}$ upper limit of 7.36\,eV for the EW of the iron K${\alpha}$ emission line. The model for the continuum emission yielded a reduced $\chi^2$ value of 1.03 for 3141 dof. The source luminosity during the 2018 {\it XMM-Newton/NuSTAR} observation, was $\approx4\times10^{36}$\,erg/s, calculated in the 0.50-30\,keV range and assuming the distance of 8.5\,kpc. All values of the best fit parameters along with errors in the 90\% confidence range are presented in Table~\ref{tab:contswift}.  

\section{Discussion}
\label{sec:discussion}

We have analyzed the high-energy spectra of {\src} and {\srcs} during different periods in their observational history. In all observations, the sources appeared to have been captured in an accretion dominated ({\it soft}) state. In this state the accretion disk  is extending close to the compact object (in our case a NS) and the observed spectrum is dominated by thermal emission. We have modeled the spectra of both sources using a combination of a multicolor disk black body and a black body model with a very high temperature (with the addition of a non-thermal tail when required). The choice for this model is motivated by the well established theoretical framework for accretion onto low-B NSs and the formation of a hot optically thick layer of material on the NS surface, known as the boundary layer \citep[e.g.,][]{1986SvAL...12..117S,2001ApJ...547..355P,2003A&A...410..217G}.  The boundary layer is a transition region between the rapidly spinning accretion disk and the NS which is rotating at a slower rate. The optically thick matter that accumulates on the NS surface is predicted to have a temperature of $>1$\,keV, producing thermal emission that accounts for a significant fraction of the total accretion luminosity.

The most notable narrow emission feature detected on top of the spectral continuum (in the energy range ${\lesssim}10$\,keV) was a prominent and broad Fe K$\alpha$ emission line, centered at $\approx6.5-6.6$\,keV. The iron emission line is observed in both sources during two different instances and in the case of \src by two different instruments. In general, the presence of iron emission is a common characteristic in the X-ray spectra of accreting compact objects  and is the result of "reflection" of X-rays from the surface of the accretion disk \citep[e.g.,~see][and references therein]{2007A&ARv..15....1D,2010LNP...794...17G}. Namely, hard X-rays are reflected  off the accretion disk, and the reflected component is registered along with the primary disk emission. The Fe K$\alpha$ line is the result of this reflection, as a fraction of high energy photons -- that are ``absorbed'' by iron atoms -- are remitted at the energy corresponding to the electron transition from the 2p orbital of the L-shell to the innermost K-shell.  While the detection of iron fluorescence in X-ray spectra of LMXBs -- such as \src and \srcs -- is a frequent and expected occurrence, a much more extraordinary event, is the fact that the emission line seems to have disappeared during some of the available observations.

\subsection{Fe K$\alpha$ line variability and the spectral state of the two UCXBs}

The spectral continuum of {\src} displays a remarkable consistency throughout all available observations, which have been carried out over two decades. Within our interpretation for the origin of the emission in \src, the bulk of the 0.5-10\,keV emission appears to be produced by an accretion disk with a temperature of ${\sim}$0.6-0.8\,keV and an inner disk radius of the order of 10\,km and an additional, hot thermal source with a ${\sim}1.5-1.8$\,keV temperature. The luminosity of the source has varied moderately by a factor of up to 25\%, and any minor variability of its spectrum is mostly manifested in the relative flux of each spectral component.  Over the course of its observational history, the spectrum of \src has been modelled using a variety of different models, including black body, disk black body models, power-law components with high energy cutoffs below 10kev, as well as broken power-laws with two spectral indexes, or more complex X-ray reflection models \citep[e.g.,~][]{2003A&A...397..249S,2003A&A...402.1021F,2011MNRAS.412L..11M,2014MNRAS.442.1157M}. We argue that the simplicity and long term consistency of our chosen model, combined with the fact that it yields physically realistic estimations of key emission parameters (see below), favors our interpretation for the origin of the source emission.
During the 1997 RXTE observation we detected a Fe K$\alpha$ emission line with an EW of ${\sim}100$\,eV, which is not present in either the Chandra or \xmm spectra with a stringent EW upper limit of 6.13\,eV and 2.86\,eV, respectively. The line is detected again in the 2017  NICER observation with an EW of ${\sim}60$\,eV.

For \srcs, we consider the same interpretation for the origin of the spectral continuum as for \src. This source also appears to have a stable spectral shape during its three recorded outbursts (in this work we present the 1st and 3rd outburst, for the 2nd outburst see \citealt{2010MNRAS.403.1426P}). However, we do note the detection of minor, in the temperatures of the thermal components; $kT_{\rm in}{\sim}0.96$\,keV in RXTE vs $kT_{\rm in}{\sim}$1.24\,keV in the {\it XMM-Newton/NuSTAR} data (the values are consistent within the 90\% range) and 2.3\,keV vs 2.8\,keV for the thermal emission (the values are consistent in the 2\,${\sigma}$ range) and a moderate variability (order of ${\times}7$ in the relative flux of the hot thermal component between observations. Nevertheless, as in the case of \src, the most striking difference between the \srcs spectrum of 2018 and the one of 2007 is the disappearance of the prominent Fe K$\alpha$ line. Namely, the bright iron emission line detected in the 2007 RXTE observation with an EW of ${\sim}$170\,eV, is not detected during the 2018 {\it XMM-Newton/NuSTAR} observation with an EW upper limit of 7.36\,eV. We also note a 25\% variation of the total source flux.

A crucial factor for establishing the reliability of our interpretation of the spectral continuum in both sources -- and more importantly for identifying the cause for the notable variability of the strength of the iron emission line -- is the rigorous estimation of true values of its spectral parameters, i.e., the temperature and inner radius of the accretion disk. To further assess the plausibility of our best-fit estimations, it is also important to compare our findings with theoretical expectations. In two of the \srcs observations (RXTE and NICER), the best value of the inner disk radius is measured in the ${\sim}$8-10.5\,km range, while for \srcs they range between ${\sim}$4 and ${\sim}$6\,km. These estimations lie below the value of the NS radius which is ${\sim}$9-15\,km depending on the equation of state of the NS \citep[e.g.,][and references therein]{2016ARA&A..54..401O}, and therefore appear to be unrealistically small. Nevertheless, we note that in all tabulated values we have not considered the necessary corrections of the thermal component parameters due the spectral hardening expected from the high plasma temperature. Correcting the inner radius estimations, assuming a moderate value of 1.8 for the spectral hardening factor (see Section~\ref{sec:hard}), yields $R_{\rm in}$ values in the ${\sim}15-45$\,km for the entire set of observations, which is the typical size for an accreting low-B NS in the soft state. 

Therefore within this framework, the presence of iron  K$\alpha$ fluorescence in the spectra of \src and \srcs is the result of illumination of the inner accretion disk by the boundary layer emission. Since the iron line is produced by X-ray reflection, its shape and strength are highly sensitive to the geometrical arrangement between the source of the primary emission (e.g.,~the boundary layer) and the accretion disk. Modification of the size and temperature of the emitting region, the inner radius of the accretion disk and the direction of the hard X-ray emission, will affect the size and location of the illuminated region of the disk. These effects will become especially pronounced in case of spectral state transition towards the so called "hard" state of accretion, were the disk recedes from the central source, giving way to the formation of an extended region of advection-dominated, optically thin radial flow. This "corona" consisting of very hot electrons produces copious amounts of non-thermal photons, resulting in a spectrum that is dominated by the power-law shaped component. Therefore, state transitions can have profound effects on the strength of the iron emission line \citep[e.g.,~][]{2010A&A...522A..96N,2010ApJ...720..205C,2014MNRAS.437..316K}. Similar effects can also be observed in accreting highly magnetized NS (i.e., high-B, X-ray pulsars), where variations in the accretion rate affect the size of the magnetoshere, the inner disk radius and the direction of the pulsar beam, resulting in pronounced variability in the iron line strength \citep[e.g.,~][]{2016MNRAS.456.3535K}. Therefore the striking variability of the iron line as observed in \src and \srcs are expected to occur in tandem with major modification of the spectral continuum and/or the source luminosity. 

However, one of the main findings of our analysis, is the notable stability of the spectrum and luminosity of both sources throughout the different observations with or without the presence of iron  K$\alpha$ fluorescence. This suggests that the attenuation of the iron line is not caused by a "macroscopic" transition in the accretion state of the source, but is likely the result of a more subtle microscopic process. To our knowledge such remarkable emission line variability in an otherwise stable X-ray binary has not been previously observed. However, the possibility of this phenomenon had been hypothesized by \cite{2013MNRAS.432.1264K} for UCXBs with C/O-rich donors.

\subsection{Variability of the Fe Ka line in UCXBs with C/O-rich donors}

In \cite{2013MNRAS.432.1264K}, it was demonstrated that the Fe K$\alpha$ emission line in the spectra of UCXBs with C/O-rich donors, is expected to be strongly attenuated, with expected values of the EW to be less than $10$\,eV. The attenuation of the iron emission line  is primarily due to the presence of overabundant oxygen\footnote{The maximum value of the O/Fe ratio corresponding to the chemical composition of a C/O white dwarf -- in which all hydrogen and helium has been converted to carbon and oxygen -- is $\approx$77 times the solar value.} in the accreting material. Namely, contrary to accretion disks with solar-like abundance, in the C/O-rich disks, absorption of photons with energies equal to or higher than the ionization threshold of iron (E$\approx7.1$\,keV), is dominated by oxygen rather than by iron. This results in the strengthening of the oxygen emission line (centered at ${\sim}0.68$\,keV) and the strong attenuation of the iron line. More specifically, after running the MCMC code of \cite{2013MNRAS.432.1264K} for an incident 1\,keV  black body spectrum  and for a range of observer viewing angles between $10^{\circ}$ and $50^{\circ}$, we find that an O/Fe ratio of at least 50 times the solar value, is required to explain the absence of a Fe K$\alpha$ line at an EW upper limit of 8\,eV. This {\it screening} effect, will hold, as long as the oxygen in the accretion disk is not fully ionized. 

The ionization state of elements in the disk was not self-consistently treated in the model developed by \cite{2013MNRAS.432.1264K}. However, the effects of the ionization state of the disk were discussed -- under a set of specific assumptions. Namely, we considered that the ionization state of the disk is determined solely by heating due to viscous dissipation as illustrated in the \cite{1973A&A....24..337S} formulation. We have ignored the effects of disk irradiation by assuming the conditions described in the \cite{2000ApJ...537..833N} analysis, predicting the formation of a thin, fully ionized layer on the disk surface which is dominated by X-ray illumination, while the temperature of deeper disk layers remain unaffected. Within this regime, \cite{2013MNRAS.432.1264K} roughly estimated that  during the {\it soft state} and for a mass accretion rate ($\dot{M}$) corresponding to $L_{X}$ of the order of a few ${\times}10^{37}$\,erg/s, oxygen could become fully ionized in the accretion disk of an UCXB, thus canceling its {\it screening} effect on the iron line. Within this framework the iron line variability in UCXBs is luminosity dependent.

More specifically assuming an optically { thick}, geometrically thin disk following the Shakura-Sunyaev ${\alpha}$-disk model, the effective temperature of the disk is given by
\begin{equation}
{T_{\rm{eff}} = {\left( {\frac{{3GM\dot M}}{{8\pi \sigma {r_{\rm{o}}}^3}}} \right)^{1/4}}{\left( {\frac{{{r\rm_o}}}{r}} \right)^{3/4}}{\left( {1 - \sqrt {{r\rm_o}/r} } \right)^{1/4}}}
\end{equation}
where $\sigma  = 5.67\cdot 10^{-5}{\rm{erg}}\,{\rm{c}}{{\rm{m}}^{ - 2}}\,{{\rm{s}}^{ - 1}}\,{{\rm{K}}^{ - 4}}$
is the Stefan-Boltzmann constant, $M$ is the mass of the accretor ($ \sim 1.4\,{M_ \odot }$ for a neutron star), $\dot M$ the mass accretion rate and ${r\rm_o}$ is the inner radius of the disk, which in the case of the soft state of a neutron star is close to its radius.
For a moderate $\dot{M}$, (corresponding to $L{\approx}2{\times}10^{37}$\,erg/s), the temperature of the inner parts of the disk { (up to 15\,$r_{g}$; where Rg is the gravitational radius $r_{g}=GM/c^{2}$)} will range between 5-8$\times10^6$\,K. In addition to heating due to viscous dissipation, if the accretion disk forms a boundary layer near the surface of the NS, it will cause a dissipation of the kinetic energy of the inner disk rings, which in turn will result in the inflation and heating of the innermost part of the disk \citep{1999AstL...25..269I}.  \cite{1999AstL...25..269I} calculate that the inner disk will be heated to $\sim10^7$\,K as the boundary layer forms. When the temperature reaches ${\sim}10^7$K -- and assuming that the disk plasma is in collisional ionization equilibrium, in a coronal approximation -- 90\%-100\% of oxygen will be fully ionized \citep{1982ApJS...48...95S}. 

The above analytical estimations demonstrate, that for typical conditions that can be met by moderately bright UCXBs, it is very likely that a considerable fraction of the inner accretion disk can reach a high enough temperature for all oxygen to become fully ionized. Although there is considerable uncertainty in the distance estimation for our two sources -- different estimations place \src anywhere between 1.4-11.5\,kpc \citep{2018ATel11302....1S,2018AJ....156...58B}, while \srcs is assumed to be at a distance of 8.5\,kpc due to its proximity to the galactic bulge -- we can robustly place both sources in the in the $10^{36}-10^{37}$\,erg/s range. From our analysis, the color-corrected, maximum disk temperature obtained by the {\tt diskbb} model, during the epochs when the line is detected is ${\sim}5{\times}10^{6}$\,K for \src and ${\sim}8{\times}10^{6}$\,K for \srcs (we note that {\tt diskbb} is only an approximation of the full \cite{1973A&A....24..337S} treatment). 

The X-ray luminosity of both sources is marginally lower than the rough threshold of
$\log(L\rm_X){\sim}37.5$ suggested by \cite{2013MNRAS.432.1264K}. Similarly the
spectroscopic estimations of the accretion disk lies below the ${\approx}1$\,KeV value for the full
ionization of 100\% of the disk oxygen, but only marginally so. If we also take into account
that the state-of-the-art Monte-Carlo simulations of X-ray reflection \citep[e.g.,][]{2010ApJ...718..695G,2011ApJ...731..131G,2013ApJ...768..146G} indicate that X-ray
irradiation can increase the disk ionization at deeper layers than the
\cite{2000ApJ...537..833N} estimations, we can reasonable assume that our sources lie
exactly at the threshold between having most of their oxygen partially ionized (suppression of
Fe K${\alpha}$ line) to fully ionized (re-appearance of Fe K${\alpha}$ line). We note that in
both cases the overabundance of oxygen expected in the C/O-rich disk (50--80 times the Solar
value) ensures the presence of notable a $O_{\rm VIII}$ fluorescent line even when most of it
becomes fully ionized (although the actual detection of such a feature depends on multiple
additional parameters; see discussion in \citealt{2013MNRAS.432.1264K}). Indeed this line is
detected at ${\sim}$0.68keV in \src, where the low absorption allows for its detection (see
Figure~\ref{fig:Chandra} and refer to \citealt{2019ApJ...883...39L} for an analysis of the NICER
observation).

We argue that the disappearance of the iron line occurring within an otherwise non-variable spectral state, is a strong indication that \src and \srcs are both C/O-rich UCBXs observed at the threshold at which the screening of the iron line ceases to exist. As a result of this precarious state, subtle variations on the parameters of the  accretion process, may lead to an increase in the fraction of fully ionized oxygen, evidenced by the reappearance of the iron emission line. Besides this striking variability in the Fe K${\alpha}$ line flux, the change in the ionization state has negligible effects on other observables.   We consider this discovery an encouraging result with respect to the \cite{2013MNRAS.432.1264K}  theoretical predictions on X-ray reflection off H-poor disks in UCXBs and their potential use as a diagnostic of the composition of the accretion disk and donor star in these sources.

\section{Summary and conclusions}
\label{sec:conclusion}
We have analyzed multiple observations of known UCXBs \src and \srcs, carried out at different eras within a 20\,year period and with different instruments. We have demonstrated that both sources (one is a persistent and the other a transient source), exhibit a remarkable stability of the shape of their spectral emission and their X-ray luminosity (during outburst). Nevertheless, we also detected a pronounced difference in the spectral characteristics at different epochs, in the form of a striking variability in the strength of the iron K${\alpha}$ emission line. Namely, the emission line is either detected as a prominent feature with an EW of 100-170\,eV, or completely disappears with an EW upper limit of less than 8\,eV. Based on the stability of the spectral shape and flux of both sources throughout the different observations, and on the values of their spectral parameters, we have attributed the iron line variability to a change in the ionization state of oxygen in the inner accretion disk regions. Namely, subtle variations of the accretion disk parameters can lead to a transition from partially to fully ionized oxygen in the disk, which in turn affects the strength of the iron line, as has been predicted for UCXBs with C/O-rich donors by \cite{2013MNRAS.432.1264K}. We argue that this behavior supports the theoretical arguments of this work and favors the C/O-rich classification of the UCXBs \src and \srcs.

\section*{Acknowledgements}
The French team are grateful to Centre National d'\'Etudes Spatiales (CNES) for their support in the work related to XMM-Newton, NuSTAR and NICER. F.K.~extends his deep gratitude to the referee and editor of this paper, for significantly contributing to its final form, but more importantly for their patience and understanding during a prolonged pause in the refereeing process, which was caused by F.K., due to an unforeseen personal issue.

\section*{Data availability}

The observational data underlying this article are publicly available at the \xmm Science Archive\footnote{http://nxsa.esac.esa.int/nxsa-web/}. Any details with regard to the Monte Carlo source code can be shared on reasonable request to the corresponding author.

\bibliographystyle{mnras}
\bibliography{general}

\label{lastpage}
\end{document}